\documentclass[twocolumn,showpacs,showkeys,amsmath,amssymb,aps,prb]{revtex4}
\usepackage{graphicx}
\usepackage{bm}

\begin{document}

\title{Quantum simulation of spin ordering with nuclear spins in a solid state lattice}

\author{Georgios Roumpos}
	\email{roumpos@stanford.edu}
	\affiliation{E. L. Ginzton Laboratory, Stanford University, Stanford, CA, 94305, USA}
\author{Cyrus P. Master}
	\affiliation{E. L. Ginzton Laboratory, Stanford University, Stanford, CA, 94305, USA}
	\affiliation{National Institute of Informatics, Hitotsubashi, Chiyoda-ku, Tokyo 101-8430, Japan}
\author{Yoshihisa Yamamoto}
	\affiliation{E. L. Ginzton Laboratory, Stanford University, Stanford, CA, 94305, USA}
	\affiliation{National Institute of Informatics, Hitotsubashi, Chiyoda-ku, Tokyo 101-8430, Japan}

\date{March 20, 2007}

\pacs{76.60.-k, 75.10.Dg, 03.67.Lx}
\keywords{Quantum simulation, Nuclear magnetism, Nuclear magnetic resonance}

\begin{abstract}
An experiment demonstrating the quantum simulation of a spin-lattice Hamiltonian is 
proposed. Dipolar interactions between nuclear spins in a solid state lattice can be
modulated by rapid radio-frequency pulses. In this way, the effective Hamiltonian of the
system can be brought to the form of an antiferromagnetic Heisenberg model with long
range interactions. Using a semiconducting material with strong optical properties such
as InP, cooling of nuclear spins could be achieved by means of optical pumping. An
additional cooling stage is provided by adiabatic demagnetization in the rotating frame
(ADRF) down to a nuclear spin temperature at which we expect a phase transition from
a paramagnetic to antiferromagnetic phase. This phase transition could be
observed by probing the magnetic susceptibility of the spin-lattice. Our calculations
suggest that employing current optical pumping technology, observation of this phase
transition is within experimental reach.
\end{abstract}

\maketitle

\section{Introduction}
Many-body problems are common in condensed matter physics.
However, computations of interesting quantities, such as critical exponents of a phase
transition, are difficult, specifically because the dimension of the associated Hilbert
space increases exponentially with the system size.
Recent quantum simulation experiments using cold atoms in an optical lattice
\cite{Greiner2002,Garcia-Ripoll2004} have opened new possibilities for the study of
these systems. The idea consists of controlling the interactions inside a many-body
system so that they take a desired form, and performing measurements on the
macroscopic behavior of the system.
It is interesting to explore how this quantum simulation concept could be implemented
in different physical systems.

Spin-lattice models are a convenient, simple model to describe magnetic phenomena
\cite{StanleyBook}. It was
recently realized, however, that various apparently unrelated problems can be cast
into this same language. An example is the half-filled Hubbard model in the limit of
large positive on-site Coulomb repulsion $U$ \cite{Auerbach}. This observation makes
spin-lattice
models interesting not only for fundamental studies of magnetism, but also as a means
to attack several other problems. Although spin-lattice models have been studied for
about a century, exact results only exist for special cases of low
spatial dimensionality. Approximation methods such as mean-field theory offer an
alternative, but they typically involve assumptions that are not rigorously justified.
On the other hand, numerical methods such as Monte Carlo simulation
\cite{LandauBook} have limited efficiency for calculations in real three-dimensional
magnetic systems. This motivates the investigation of other techniques to study
spin-lattice models.

In this paper, we consider a solid-state lattice of nuclear spins as a simulator of a
specific spin-lattice model, in which interactions between spins are modulated by rapid
radio-frequency pulses. The potential of nuclear magnetic resonance (NMR) techniques
\cite{Ramanathan2004,AbrBook} for simulating an artificial many-body Hamiltonian is
theoretically explored. Nuclear spins interact predominantly via a magnetic dipolar
interaction, which can be modulated by varying the orientation of the crystal relative to
the applied magnetic field, applying NMR pulse sequences, or using different material
systems.
By optically cooling the nuclear spins \cite{Meier1984,Tycko1996}, or by implementing a
similar dynamic polarization technique, it is possible that the system eventually reaches
a nuclear spin temperature \cite{AbrBook,Abr1958,GoldBook} where a phase transition
\cite{StanleyBook} occurs. In particular, in Sec. \ref{sec:ExpSyst} we describe the
transition from a paramagnetic to an antiferromagnetic phase that we expect to occur
in our proposed system.
In this context, we are not considering a generic spin-Hamiltonian simulator, but
rather a problem-specific machine, where we have the ability to change some aspects of
the spin-lattice Hamiltonian and the spin temperature, using the freedom provided by
the particular experimental system.

The proposed experiment proceeds as follows: We first optically pump a sample of bulk
InP in low physical temperature and high magnetic field, so that the resulting
nuclear-spin polarization is maximum \cite{Michal1999, Goto2004}.
We then perform adiabatic demagnetization in the rotating frame (ADRF) to transfer the
Zeeman order to the dipolar reservoir \cite{SlichterBook,Oja1997}.
This is a cooling technique analogous to slowly switching off the dc magnetic field in
the laboratory frame. If it is performed adiabatically, i.e. with no loss of entropy, the
ordering due to alignment of the spins along the dc magnetic field is transferred to
ordering due to local correlations of the spin orientations.
By applying a suitably chosen NMR pulse sequence, we can transform the inherent
Hamiltonian of the system to the desired one in an average sense, to be described
below. In particular, we use NMR techniques to engineer an
antiferromagnetic Heisenberg model with long range interactions.
This transformation is validated by average Hamiltonian theory (AHT) \cite{Haeberlen}.
Our system has two species of nuclear magnetic moments. We argue that in the
ground state of the described Hamiltonian each nuclear species is ordered
antiferromagnetically, so that each nuclear species consists of two sublattices
of opposite spin orientation.
Antiferromagnetic ordering with two sublattices can be observed in the NMR spectrum
as the appearance of two peaks of opposite phase: In a system with two nuclear
species A and B of widely-separated Larmor frequencies, we can rotate the spins
A to the horizontal plane using a $\frac{\pi}{2}$-pulse, and observe their free
induction decay (FID) signal. If the spins B are in an antiferromagnetically ordered
state, they create a spatially oscillating local field that modulates the Larmor
frequency of spins A. This means that spins A in different sublattices will produce
the FID signals of different frequencies. However, to decide whether a phase
transition occurs, we need to measure the magnetic susceptibility of the system
\cite{Gold1974,Mouritsen1980}. In particular, the longitudinal susceptibility has
a maximum at the transition temperature, while the transverse susceptibility
exhibits a plateau inside the ordered phase. Nuclear spin susceptibilities can
be easily extracted from the Fourier-transformed FID signals recorded after
a short pulse. This measurement exploits the Fourier-transform relationship
between transient and steady-state response of a system to a small perturbation.

All individual pieces of this experiment have already been demonstrated, but putting
them together is challenging. Nuclear ordering has been observed in a series of
experiments \cite{Oja1997,Chapellier,Quiroga1,Quiroga2}.
However, the effect of NMR pulse sequences to engineer an artificial Hamiltonian has
not been studied in detail, and this is an issue that the current study addresses.
We should note that the class of Hamiltonians that can be simulated with this method
is severely limited by the characteristics of the particular material system, although
many interesting cases can be studied, for example frustrated spin lattices.

This paper is organized as follows: In Sec. \ref{sec:ADRF}, we briefly describe
adiabatic demagnetization in the rotating frame (ADRF), which is essentially a cooling
technique for nuclear spins.
Sec. \ref{sec:calculation} discusses the calculation of thermodynamic quantities for
systems obeying the spin Hamiltonians of interest, and in particular, the evolution
of spin temperature during ADRF in our proposed material system.
In this way, we can calculate the spin temperature reached by the spin-lattice after
ADRF. In the subsequent two sections, we describe the spin Hamiltonians that can be
simulated by this method.
In particular, a suitable NMR pulse sequence is discussed in Sec.
\ref{sec:PulseSeq}, and an experimental material system is presented in Sec.
\ref{sec:ExpSyst}.
Finally, in Sec. \ref{sec:MeanField}, we give an estimate, based on mean-field
approximation, of the initial polarization needed to observe a phase transition.

\section{\label{sec:ADRF}Description of adiabatic demagnetization in the rotating
frame (ADRF)}
The term {\it adiabatic} is used in two different meanings in physics, one quantum
mechanical, and the other thermodynamic \cite{AbrBook,Abr1958}. ADRF is based on
the thermodynamic definition, which is a reversible process with no heat transfer, in
which the entropy of the system remains constant. The term {\it isentropic} also
applies to this case. By contrast, in quantum mechanics, the term {\it adiabatic}
usually means that the relative populations of the various energy eigenstates of the
system are kept constant. The two definitions are in general incompatible.

The system of nuclear spins interacts only weakly with the crystal lattice. The relevant
timescale for this interaction is $T_1$ \cite{AbrBook}. So, if we are interested in
phenomena much faster than $T_1$, we can treat the system as being
effectively isolated, and describe it in terms of its nuclear spin temperature
\cite{GoldBook}, which can be very different from the lattice temperature.
Of course, if we let the system to relax over several timescales $T_1$, these two
temperatures eventually become equal.

Consider a lattice of nuclear spins in a dc Zeeman magnetic field $B_0$ along the
$z$-direction. During ADRF, we apply a rotating radio-frequency (rf) magnetic field in
the $x$-$y$ plane of magnitude $B_1$ and angular frequency $\omega_1$, so that the
Hamiltonian in the frame rotating with angular frequency $\omega_1$ (rotating frame) is
\begin{equation}
\mathcal{\hat{H}}^{rot} = -\sum_i \hbar\left[ \left(\gamma_i B_0 - \omega_1 \right)\hat{I}_i^z
+ \gamma_i B_1\hat{I}_i^x \right] +\mathcal{\hat{H}}^\prime_D,
\label{eq:totHamiltonian}
\end{equation}
where $\hat{I}^{x,y,z}_i$ and $\gamma_i$ are the spin operators and gyromagnetic ratio
of the $i^\text{th}$ spin dipole, and $\mathcal{\hat{H}}^\prime_D$ denotes the secular
part of the dipolar interaction Hamiltonian. The latter follows after dropping the rapidly
oscillating terms. The three terms in the above equation we called the Zeeman, the rf
(radio frequency), and the dipolar part of the total Hamiltonian, respectively. In the
case where the lattice consists of nuclear spins of the same species
\cite{SlichterBook}
\begin{align}
\mathcal{\hat{H}}^\prime_D &= \frac{\mu_0}{4\pi} \hbar^2 \gamma_I^2 \frac{1}{2}
\sum_{i\neq j}\frac{1-3\cos^2(\theta_{ij})}{r_{ij}^3}
\left( \hat{I}_i^z \hat{I}_j^z -\frac{1}{2}\hat{I}_i^+\hat{I}_j^- \right) \nonumber \\
& \equiv \frac{1}{2} \sum_{i, j}u_{ij}
\left( \hat{I}_i^z \hat{I}_j^z -\frac{1}{2}\hat{I}_i^+\hat{I}_j^- \right) = \mathcal{\hat{H}}^\prime_{II} .
\label{eq:dipHamRotSameSpec}
\end{align}
The distance between the spins $i$ and $j$ is $r_{ij}$, $\theta_{ij}$ is the angle
between the vector $\bm{r}_{ij}$ and the direction of the Zeeman magnetic field
$\hat{z}$, while $\hat{I}_i^\pm =\hat{I}_i^x\pm i \hat{I}_i^y$ and $\gamma_I$ is the
common gyromagnetic ratio of the spins \footnote{We have also set $u_{ii}=0$ to get
rid of the restriction $i \neq j$ in the sum.}.
Note the role of the term $-\sum_i \hbar \gamma_i B_1\hat{I}_i^x$ in
(\ref{eq:totHamiltonian}) allowing the Zeeman and dipolar reservoirs to exchange
energy. In the absence of the rf field, the Zeeman and dipolar parts of the
Hamiltonian are separate constants of motion, and the conditions of thermal
equilibrium are not satisfied. In the presence of the rf field, a common
spin-temperature is established, and the system is described in the rotating frame by
the density matrix
\begin{equation}
\hat{\rho}^{rot} =
\frac{\exp\left( -\beta \mathcal{\hat{H}}^{rot} \right)}{\mathrm{Tr}
\left[ \exp\left( -\beta \mathcal{\hat{H}}^{rot} \right) \right]}.
\label{eq:rhoRotGen}
\end{equation}
The process of establishment of thermal equilibrium is described by
the Provotorov equations \cite{AbrGold82}.
When the rf term becomes comparable to the Zeeman term, Zeeman order is 
established along the direction of the effective magnetic field
$\left( B_0-\frac{\omega_1}{\gamma_I}\right)\hat{z} + B_1\hat{x}$.
After $\omega_1$ reaches $\gamma_I B_0$, we switch off the rf field $B_1$
adiabatically, so that all the initial Zeeman order is transferred to the dipolar
reservoir.

\section{\label{sec:calculation}Evolution of spin-temperature during ADRF}
During ADRF, the entropy of the system remains constant. So, if we have an
expression for the entropy as a function of the inverse spin-temperature $\beta$,
we can calculate the spin-temperature during the ADRF.
In particular, we can estimate the desired initial polarization, that is required
in order to observe a phase transition.

In the rest of the paper, we drop the superscript ${rot}$, as we are only going to
work in the rotating frame. We also use the word \emph{temperature} to denote
\emph{nuclear spin-temperature}.

In the high temperature approximation \cite{SlichterBook}, we can expand the
expression for the density matrix (\ref{eq:rhoRotGen}) to first order in inverse
temperature $\beta$
\begin{equation}
\hat{\rho} = A\left(\hat{1}-\beta \mathcal{\hat{H}}\right).
\label{eq:rhoLinExp}
\end{equation}
where $A$ is a suitably chosen constant, so that $\rm{Tr}(\hat{\rho}) = 1$.
The entropy $\mathcal{S}$ then follows from the equation
\begin{equation}
\mathcal{S} = -k_B \mathrm{Tr}\left(\hat{\rho} \ln \hat{\rho}\right).
\end{equation}
For $N$ spin-$j$ particles of the same species, with gyromagnetic ratio $\gamma$,
interacting through the dipolar interaction of eq. (\ref{eq:dipHamRotSameSpec})
we find, in the rotating frame
\begin{equation}
\frac{\mathcal{S}}{N k_B} =
\ln (2j+1) -\beta^2\hbar^2\gamma^2\frac{j(j+1)}{6}
\left[\left(B_0- \frac{\omega_1}{\gamma}\right)^2 +B_L^2\right],
\label{eq:EntropyLocalField}
\end{equation}
where
\begin{equation}
B_L^2 \equiv \frac{3}{4} \frac{j(j+1)}{3\hbar^2\gamma^2} \sum_j\left(u_{ij}\right)^2.
\label{eq:LocalField}
\end{equation}
In the above equation, we have ignored the rf field $B_1$.
$B_L$ can be interpreted as a ``local field'' due to the neighboring spins.

We follow the procedure described in Refs.~\onlinecite{Gold75,SHEB62,AbrGold82}
(see also Ref.~\onlinecite{Izyumov1988} for similar calculations)
to extend this description to the case of lower temperature. The idea is to include
more terms in the expansion (\ref{eq:rhoLinExp}), applying a diagrammatic
technique similar to that used in perturbation theory. The Zeeman term is
considered as the free part of the Hamiltonian, and is treated to all orders.
The dipolar interaction term is considered as the perturbation, and by
grouping the resulting terms using a procedure that recalls Wick's
theorem \cite{Wick1950}, we are able to efficiently perform the summations.
In this calculation, we ignore the rf term in (\ref{eq:totHamiltonian}), as this has
only the effect of bringing the Zeeman and dipolar reservoirs to equilibrium.
Also, far from resonance, the two spin species cannot exchange energy, because their
interaction takes the form $\sum_{i k}v_{ik}  \hat{I}_i^z \hat{S}_k^z$, which cannot
induce flip-flops. When the rf part in equation (\ref{eq:totHamiltonian}) becomes
comparable with the Zeeman part, the two reservoirs can exchange energy, because
in this case the Zeeman ordering is not along the $\hat{z}$ direction.
This means that the Zeeman and interaction parts of the Hamiltonian do not commute,
so they are not separately constants of motion, and the conditions for the
establishment of thermal equilibrium are satisfied.
In our approximation, we do not study the effect of different spin-temperatures.
Our calculation consists of defining the initial polarization of one spin species, and
calculating the spin-temperature in the rotating frame during the ADRF, with the
constraints that the system is described by one spin-temperature, and that the
entropy is invariant.

To study the feasibility of observing a phase transition, we estimate $T_c$ by the
condition that the longitudinal magnetic susceptibility vanishes, as the ordered state
is expected to be an antiferromagnet.
This is not an accurate calculation, however, because only a few diagrams are
involved in the high-temperature expansion, while in the critical region higher order terms
become increasingly important. A common procedure to take the latter into account
is described in Ref. \onlinecite{StanleyBook}, in which extrapolation procedures
are used to compute the limiting behavior of high-temperature expansion series given
only the knowledge of the first few terms.
Also, antiferromagnets have a more rich phenomenology than this simple picture. In
particular, Refs.~\onlinecite{Haldane1-1983,Haldane2-1983,Auerbach} imply that 1D
Heisenberg antiferromagnetic spin chains, in the limit of large spin, are gapless for
half-integer spin, so that the susceptibility does not vanish at the critical point.
The mean-field description \cite{Gold1974} is quite different, as well.
A much better way, based on mean-field theory, to calculate critical quantities, is
presented in Sec. \ref{sec:MeanField}.

\subsection{\label{sec:LongitudinalCase}The longitudinal case}
Assume that the spin interaction is longitudinal; that is, the Hamiltonian has the form
\begin{equation}
\mathcal{\hat{H}} = \Delta \sum_i \hat{I}_i^z + \frac{1}{2}\sum_{i,j}u_{ij}\hat{I}_i^z \hat{I}_j^z \equiv
\mathcal{\hat{H}}_Z + \mathcal{\hat{H}}^\prime.
\end{equation}
$\Delta$ stands for the detuning $\hbar \left( \gamma_I B_0 -\omega_1 \right)$.
After thermalization, the system is described by the density matrix
\begin{equation}
\hat{\rho} =
\frac{\exp\left(-\beta_0\mathcal{\hat{H}}_Z-\beta\mathcal{\hat{H}}^\prime\right)}{\mathrm{Tr}
\left[ \exp\left(-\beta_0\mathcal{\hat{H}}_Z-\beta\mathcal{\hat{H}}^\prime\right) \right]}.
\end{equation}
We allow for different temperatures for the Zeeman and dipolar reservoir for
convenience in the calculation. In the end, we are going to set these temperatures
equal. Defining the
zeroth order thermal average for an operator $\mathcal{\hat{Q}}$ as
\begin{equation}
\langle \mathcal{\hat{Q}} \rangle_0 =
\frac{\mathrm{Tr}\left[\exp\left(-\beta_0\mathcal{\hat{H}}_Z\right)
\mathcal{\hat{Q}}\right]}{\mathrm{Tr}
\left[ \exp\left(-\beta_0\mathcal{\hat{H}}_Z\right) \right]},
\end{equation}
and treating the interaction term as a perturbation, we have the following expansion
for the true thermal average
\begin{equation}
\langle \mathcal{\hat{Q}} \rangle = \frac{\sum_{n=0}^\infty \left(-\beta\right)^n
\langle\left(\mathcal{\hat{H}}^\prime\right)^n \mathcal{\hat{Q}}\rangle_0 /n! }{
\sum_{n=0}^\infty \left(-\beta\right)^n \langle\left(\mathcal{\hat{H}}^\prime
\right)^n\rangle_0 /n!}.
\label{eq:QSeriesExp}
\end{equation}
$\left(\mathcal{\hat{H}}^\prime\right)^n$ involves multiple sums, and the bookkeeping
becomes complicated, unless we define the semi-invariants $M^0_i$
\begin{equation}
\langle (\hat{I}_j^z)^n \rangle_0 = \sum_{i_1+i_2+\ldots+i_p=n} M^0_{i_1} M^0_{i_2} \ldots
M^0_{i_n}.
\label{eq:SemiInvDef}
\end{equation}
With this definition, we can write down a diagrammatic expansion, like the one
described in Ref.~\onlinecite{SHEB62}. The advantage over the brute-force expansion
(\ref{eq:QSeriesExp}) is that the resulting sums are unrestricted, i.e. no constraints
such as $i \neq j$ are involved in the summations.
If we define connected diagrams as those involving $\mathcal{\hat{Q}}$, and
disconnected ones as those not involving it, we can factor the numerator as the
product of the sum of connected diagrams times the sum of disconnected ones.
The latter serves to cancel the denominator, so that only connected diagrams matter.

The calculation of the semi-invariants is straightforward. With the definitions
\begin{align}
\label{eq:fjDef}
\eta & = -\beta_0 \Delta, \nonumber \\
f_j(\eta) & = \mathrm{tr}\left( e^{\eta \hat{I}^z} \right) = 
\frac{\sinh\left[\eta \left(j+\frac{1}{2}\right) \right]}{\sinh\left(\frac{\eta}{2}\right)}, \\
t_j^{(n)} & = \frac{1}{f_j(\eta)}\left( \frac{\mathrm{d}}{\mathrm{d}\eta} \right)^n
f_j(\eta), \nonumber
\end{align}
the first few semi-invariants can be written as
\begin{subequations}
\begin{align}
M^0_1 & = t_j^{(1)},\\
M^0_2 & = t_j^{(2)} - \left( t_j^{(1)} \right)^2, \\
M^0_3 & = t_j^{(3)} - 3 t_j^{(2)} t_j^{(1)} + 2 \left( t_j^{(1)} \right)^3.
\end{align}
\end{subequations}
The general case is covered in Appendix \ref{sec:GeneralCase}

\subsection{\label{sec:CalcThermQuant}Calculation of thermodynamic quantities}
In this subsection, we calculate the entropy as a function of inverse temperatures
$\beta$ and $\beta_0$ and detuning $\Delta$. In the final formulas, we always set
$\beta_0 = \beta$, but we consider $\beta$ and $\beta_0$ as independent variables
in the calculation of derivatives. We first need to calculate the average polarization.
\begin{equation}
\Delta \frac{\partial}{\partial \beta}\langle \hat{I}^z \rangle =
\frac{\partial}{\partial \beta_0}\langle \mathcal{\hat{H}}^\prime_D \rangle =
\frac{\partial}{\partial \eta}\langle \mathcal{\hat{H}}^\prime_D \rangle
\frac{d \eta}{d \beta_0} =
-\Delta \frac{\partial}{\partial \eta}\langle \mathcal{\hat{H}}^\prime_D \rangle.
\end{equation}
The dipolar interaction energy $\langle \mathcal{\hat{H}}^\prime_D \rangle$ is
calculated in Appendix \ref{sec:CalcDipolarInt} in terms of the functions
$G_n(.)$ defined there. Performing the integral in the above equation, we find
\begin{multline}
\frac{\langle \hat{I}^z\rangle(\beta_0,\beta)}{N} = t^{(1)} \\
-\sum_n \frac{\beta^{n+1}}{n+1}
\frac{\partial}{\partial \eta}G_n\left(t^{(1)}(\eta),t^{(2)}(\eta),
\ldots,t^{(n+1)}(\eta)\right).
\end{multline}
If we keep terms to first order in $\Delta$, we can estimate the longitudinal
magnetic susceptibility.
\begin{multline}
\frac{1}{N} \langle \hat{I}^z \rangle \simeq \\
-(\beta \Delta) \frac{j(j+1)}{540} \left[180-(33+56j+56j^2)\beta^2
\left(\gamma\hbar B_L\right)^2\right].
\end{multline}
This formula provides a method to estimate the critical temperature of the
paramagnetic to antiferromagnetic phase transition from the temperature at which
 the susceptibility becomes zero. For spin-$\frac{1}{2}$ and spin-$\frac{9}{2}$, we find
\begin{subequations}
\begin{equation}
k_B T_C (j=\frac{1}{2}) = 0.645 \left(\gamma\hbar B_L\right),
\end{equation}
\begin{equation}
k_B T_C (j=\frac{9}{2}) = 2.81 \left(\gamma\hbar B_L\right).
\end{equation}
\label{eq:TcSusceptibility}
\end{subequations}
This is a very rough estimate, as discussed in the beginning of this section. Further,
it refers to the ordering of each  species in its own lattice, not involving coupling
between them. Eventually we will use these couplings to create the artificial
Hamiltonian described in Sec. \ref{sec:PulseSeq}.

The free energy follows from the expression
\begin{equation}
\ln \mathcal{Z}(\beta_0,\beta) = \ln \mathcal{Z}(\beta_0,0)
- \int_0^\beta \langle \mathcal{\hat{H}}^\prime_D \rangle (\beta_0,\lambda) d\lambda,
\end{equation}
which can be expressed in terms of the functions $G_n$
\begin{multline}
\frac{1}{N} \ln \mathcal{Z}(\beta_0,\beta) = \ln f_j(-\beta_0 \omega_0) \\
- \sum_n \frac{\beta^{n+1}}{n+1}
G_n\left(t^{(1)}(\eta),t^{(2)}(\eta),\ldots,t^{(n+1)}(\eta)\right).
\end{multline}

We are now ready to calculate the entropy.
\begin{equation}
\frac{\mathcal{S}}{k_B} = \beta_0 \omega_0 \langle \hat{I}^z \rangle
+ \beta \langle \mathcal{\hat{H}}^\prime_D \rangle + \ln \mathcal{Z}(\beta_0,\beta);
\end{equation}
in terms of the functions $G_n$,
\begin{equation}
\frac{\mathcal{S}}{N k_B} = -\eta t_j^{(1)} + \ln f_j(\eta)
+ \sum_n \frac{\beta^{n+1}}{n+1}
\left( nG_n + \eta \frac{\partial}{\partial \eta}G_n \right).
\end{equation}

\subsection{\label{sec:results}Results}
During ADRF, the entropy remains constant. In order to obtain an ordered final state,
we need the initial state to be sufficiently polarized. A rough estimate for
spin-$\frac{1}{2}$ species is to set the spin temperature $k_BT$ of the initial
paramagnetic state equal to the level spacing $\hbar\gamma B_0$. This means that we
need the initial polarization, which is defined as the thermal average magnetization
over maximum magnetization, to be
\begin{equation}
p = \tanh\left( \frac{\hbar\gamma B_0}{2k_BT} \right) = 0.46.
\end{equation}
Sec. \ref{sec:MeanField} gives a better picture of how the initial polarization
influences the ordering observed under the artificial Hamiltonian we wish to implement.

\begin{figure}
\includegraphics[width=2.8in]{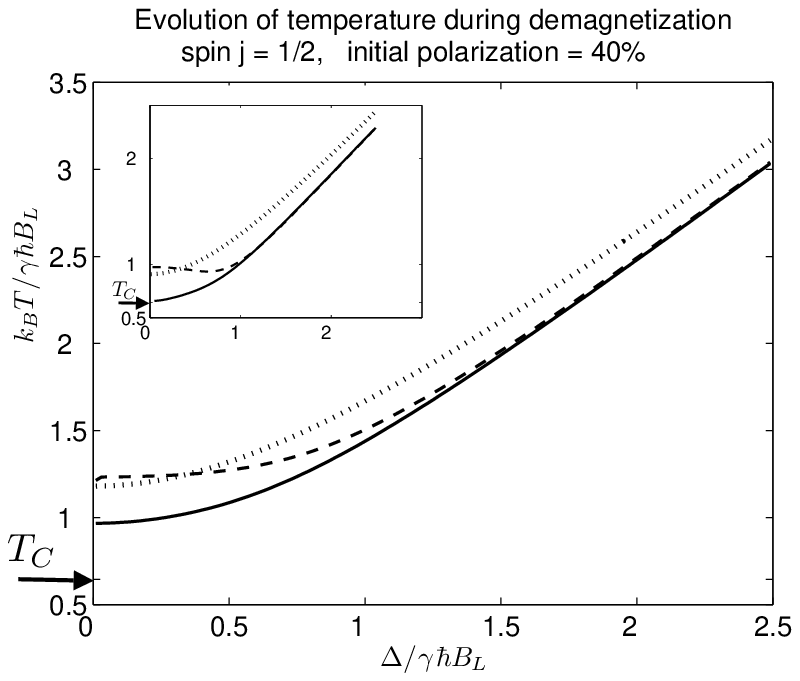}
\includegraphics[width=2.8in]{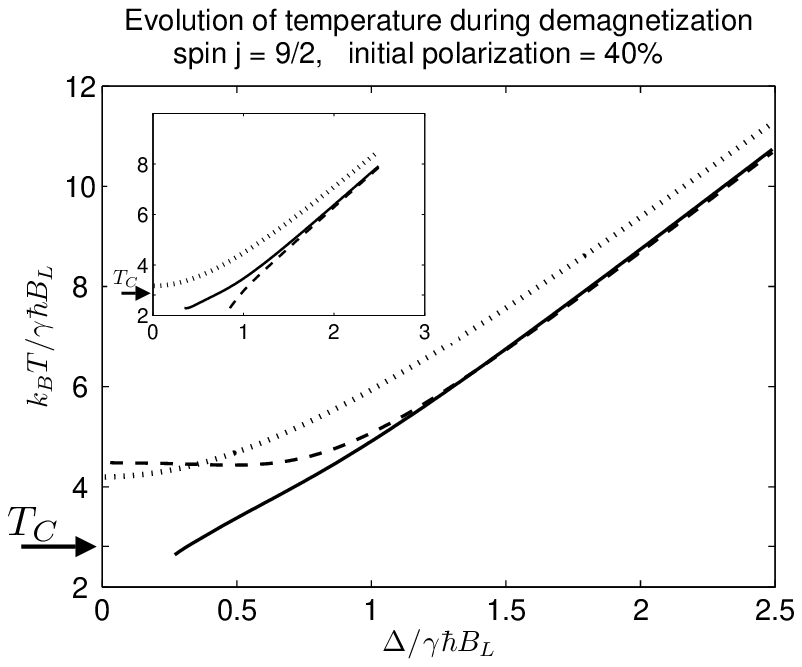}
\caption{\label{fig:results}Evolution of spin temperature $T$ vs detuning $\Delta$
during ADRF. Dotted line: high temperature approximation. Broken line: first order
approximation. Continuous line: second order approximation. These approximations
are discussed in Sec. \ref{sec:calculation} and in the Appendices. Insets: Initial
polarization of 50\%. The parameter $B_L$ (local field) is defined in
(\ref{eq:LocalField}), $T_c$ is calculated in (\ref{eq:TcSusceptibility}).
The calculation involves one spin species with spin-$\frac{1}{2}$ or $\frac{9}{2}$ in
an fcc lattice and the Zeeman magnetic field along the (111) crystalline direction.
For high initial polarization, large spin, and near
resonance, the calculation does not converge.}
\end{figure}

In Fig. \ref{fig:results} we present the results of our numerical calculation.
About 50\% initial polarization of the spins is required in order to
reach a spin temperature where a phase transition can be observed.
Details about the proposed material system are presented in Sec. \ref{sec:ExpSyst}
and in the Appendices.

\section{\label{sec:PulseSeq}Pulse sequence and the artificial Hamiltonian}
If the system has been sufficiently cooled via optical pumping and ADRF,
it might be in dipolar-ordered state because of the inherent interaction
between spins. On the other hand, by applying a
WAHUHA\cite{WAHUHA1,WAHUHA2}, or
MREV8\cite{MREV8-1,MREV8-2,MREV8-3}
pulse sequence\cite{Haeberlen} on resonance, we
can decouple spins of the same species, and transform the interaction between unlike
spins from an Ising-like interaction to a Heisenberg-like interaction.

In particular, the WAHUHA pulse cycle is given by
\begin{equation}
\tau -P_x -\tau -P_{-{y}} - 2\tau - P_y - \tau - P_{-{x}} - \tau,
\end{equation}
where $P_\alpha$ and $P_{-\alpha}$ denote $\frac{\pi}{2}$ and $-\frac{\pi}{2}$ pulses
respectively around the $\alpha$-axis, ($\alpha=x,y$). $\tau$ and $2\tau$ denote the
time interval between the application of consecutive pulses. The pulse cycle time is
$t_c = 6 \tau$ plus the duration of the four pulses.
MREV8 contains two WAHUHA-type pulse cycles, concatenated to mitigate
pulse imperfection effects.
If the pulses are short enough, then their effect can be modeled as an instantaneous
transformation of the Hamiltonian. To first order in $t_c$ relative to the magnitude
of the Hamiltonian, the system dynamics are described by the time-averaged
Hamiltonian over the pulse cycle. Average Hamiltonian theory (AHT) is covered in detail
in several textbooks (see, for example, Ref.~\onlinecite{Haeberlen}).
To first order in AHT, the dipolar interaction in the rotating frame between nuclei of
the same species (\ref{eq:dipHamRotSameSpec}) is eliminated. If we simultaneously
apply this pulse sequence to both species, their dipolar interaction in the rotating
frame
\begin{equation}
\mathcal{\hat{H}}^\prime_{IS} = \sum_{i,k}w_{ik}\hat{I}_i^z \hat{S}_k^z
\end{equation}
is transformed to a Heisenberg-like interaction
\begin{equation}
\hat{\overline{\mathcal{H}^\prime}}_{IS} =
\sum_{i,k}w_{ik}\frac{1}{3}\hat{\bm{I}}_i \cdot \hat{\bm{S}}_k.
\label{eq:IntHeis}
\end{equation}
In the above sums, the index $i$ runs over the spins $I$, whereas the index $k$ runs
over the spins $S$. $w_{ik}$ has a similar expression as $u_{ij}$ in
(\ref{eq:dipHamRotSameSpec}).
\begin{equation}
w_{ik} = \frac{\mu_0}{4\pi}\hbar^2\gamma_I\gamma_S
\frac{1-3\cos^2(\theta_{ik})}{r_{ik}^3}.
\end{equation}

If we use the modified WAHUHA pulse cycle
\begin{equation}
\tau_1 -P_x -\tau_2 -P_{-{y}} - 2\tau_2 - P_y - \tau_2 - P_{-{x}} - \tau_1,
\end{equation}
the effective interaction Hamiltonians to first order in AHT are
\begin{equation}
\hat{\overline{\mathcal{H}^\prime}}_{II}(\tau_1, \tau_2) = \frac{\tau_1-\tau_2}{t_c}
\sum_{i,j} u_{ij}\left( \hat{I}_i^z \hat{I}_j^z -\frac{1}{2}\hat{I}_i^+\hat{I}_j^- \right) 
\end{equation}
\begin{equation}
\hat{\overline{\mathcal{H}^\prime}}_{IS}(\tau_1, \tau_2) =
\sum_{i,k}w_{ik} \left[ \frac{2\tau_1}{t_c}\hat{I}_i^z\hat{S}_k^z
+ \frac{2\tau_2}{t_c}\left( \hat{I}_i^x\hat{S}_k^x + \hat{I}_i^y\hat{S}_k^y\right)\right]
\end{equation}
where $t_c = 2\tau_1 + 4\tau_2$. By keeping $t_c$ constant and varying $\tau_2$
adiabatically from $0$ to $\frac{t_c}{6}$, we can get an interpolation between the initial
and final Hamiltonians $\mathcal{\hat{H}}^{init}$, $\mathcal{\hat{H}}^{fin}$, and transform
the state of the system from the thermal state of $\mathcal{\hat{H}}^{init}$ with
entropy $\mathcal{S}$ to the thermal state of $\mathcal{\hat{H}}^{fin}$ with the
same entropy. In this way, we can study the thermodynamics of the particular
Heisenberg model for varying entropy. An alternative method is to keep $\tau_1=\tau_2$
and vary the width of the pulses, so that they induce rotations by an angle $\theta$,
same for all pulses. By varying $\theta$ from 0 to $\frac{\pi}{2}$, we have a similar effect,
but the equations for the effective interactions are more complicated.

\section{\label{sec:ExpSyst}Choice of the experimental system}
\begin{figure}
\includegraphics[width=3.7in]{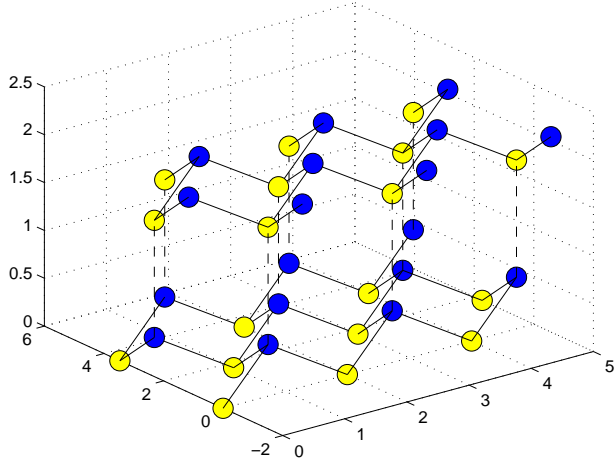}
\caption{\label{fig:planes}(color online) The configuration of the nuclear spins under
the experimental conditions described in Sec. \ref{sec:ExpSyst}. Spheres of the
same color represent nuclei of the same species. Solid lines are antiferromagnetic
bonds, while broken lines are ferromagnetic ones. Only nearest-neighbor bonds are
plotted. The perpendicular direction in this figure is the (111) crystalline
direction. We can see the postulated antiferromagnetic planes, which are coupled
ferromagnetically. Within each plane, spins of one species tend to form a
ferromagnetic lattice which is aligned antiferromagnetically to the lattice formed by
the spins of the other species. On the other hand, spins of the same species
belonging to successive planes tend to align antiferromagnetically.}
\end{figure}

Bulk InP is a suitable material for this experiment. As described in Sec.
\ref{sec:PulseSeq}, two spin species are required to implement the desired model
Hamiltonian. A strained lattice, on the other hand, like a quantum well exhibits strong
quadrupolar broadening.
This is because spins higher than $\frac{1}{2}$, such as $^{113}$In and $^{115}$In,
include a quadrupole term in the Hamiltonian, which gives a non-zero contribution in
the presence of electric field gradients. When this contribution is the dominant one,
it can be expected that any ordering due to the dipolar interaction term is destroyed.
Phosphorous has only one isotope $^{31}$P with spin-$\frac{1}{2}$, while Indium
has two isotopes, $^{113}$In and $^{115}$In have the same spin ($\frac{9}{2}$)
and are close in Larmor frequency for typical NMR magnetic fields,
so we can consider Indium to be homonuclear. InP is a direct-gap semiconductor
with strong optical properties, and efficient optical cooling has already been
demonstrated\cite{Michal1999,Goto2004}. InP has a zincblende structure; if a dc
magnetic field is applied along the (111) crystalline direction, and couplings between
spins of the same species are neglected during the application of the NMR pulse
sequence as described in Sec. \ref{sec:PulseSeq}, then the magnetic dipolar
interaction between nuclei creates the following structure (note that the gyromagnetic
ratio $\gamma$ of both In and P nuclei is positive). The planes of Indium and
Phosphorous nuclei are grouped into pairs. The coupling between nearest
neighbors (nn), which are always of different species, is antiferromagnetic if both
nuclei belong to the same pair of planes,
and ferromagnetic if the nuclei belong to successive pairs of planes.
Second-nearest-neighbors are of the same spin species, and so their interaction
is averaged out. Some third-nearest-neighbors interactions cause frustration, but
the magnetic field orientation ensures that their effect is small. Ignoring
interactions farther than third-nearest-neighbors, this leads the rather
nice picture shown in Fig. \ref{fig:planes}. Interactions within each pair
of planes are antiferromagnetic. Coupling between plane pairs is ferromagnetic. It is
possible (but by no means certain) that in the absence of an offset dc magnetic field
in the rotating frame, each plane pair acts as a two-dimensional antiferromagnet,
where the ferromagnetic interaction between plane pairs merely ensures that all of
them have the same structure. A complementary picture results if we consider
each spin species independently. In our proposed structure, each species forms
a lattice with antiferromagnetically ordered spins. Antiferromagnetic ordering
within spins of the same species results in an indirect way through interactions
with spins of the other species, as shown in Fig.  \ref{fig:planes}.

How reasonable is it for us to neglect interactions beyond 3rd nn? The strength of
dipolar interactions falls off with distance only as $\frac{1}{r^3}$, and because the
volume integral
\begin{equation}
\iiint \mathrm{d}V \frac{1}{r^3} = 4\pi \int_0^\infty \mathrm{d}r \frac{1}{r}
\end{equation}
is logarithmically divergent, interaction with distant nuclei, could significantly modify
the above picture.
We discuss this question in the next section, in which we take into account coupling
between every pair of spins within the mean-field approximation.

\section{\label{sec:MeanField}Mean field theory of ordering}
In Sec. \ref{sec:calculation} we presented an estimate of the initial nuclear spin
polarization needed in order to observe a phase transition from a paramagnetic to
an antiferromagnetic phase. Our discussion involved thermodynamic arguments.
A more exact estimate could not be obtained with this method, as our calculation
includes only a few diagrams, and is not valid near the phase transition.
Additionally,It does not converge to an answer when the spin temperature is small,
and the calculation of $T_C$ is not accurate, as already explained in the beginning
of that Sec. \ref{sec:calculation}.
We now use the local Weiss field approximation to study the problem of ordering of
nuclear spins under the effective interaction of our model. Our discussion closely
follows Refs.~\onlinecite{Gold1974,AbrGold82}.

During the application of the
decoupling pulse sequence, the spin Hamiltonian in the rotating frame is
\begin{equation}
\mathcal{H}= c \gamma_I \gamma_S \sum_{i,k}A_{ik}\hat{\bm{I}}_i \cdot
\hat{\bm{S}}_k,
\end{equation}
where $c=\frac{1}{3} \frac{\mu_0}{4\pi} \hbar^2$,
$A_{ik}=\frac{1-3\cos^2(\theta_{ik})}{\left| \bm{r}_i-\bm{r}_k \right|^3}$. Neglecting
short-range correlations between spins, we can consider every spin as being subject to
a local field
\begin{eqnarray}
\hbar \boldsymbol{\omega}_i &=& c\gamma_I\gamma_S \sum_{k}A_{ik}
\langle \hat{\bm{S}}_k \rangle , \\
\hbar \boldsymbol{\omega}_k &=& c\gamma_I\gamma_S \sum_{i}A_{ik}
\langle \hat{\bm{I}}_i \rangle .
\end{eqnarray}
We then assume thermodynamic equilibrium, so that the expectation values
$\langle \hat{\bm{I}}_i \rangle$, and $\langle \hat{\bm{S}}_k \rangle$ equal their thermal
averages. Near a phase transition, the magnetization is small, so that we can neglect
all terms except the linear term
\begin{equation}
\langle \hat{\bm{I}}_i \rangle = -\frac{1}{3} I(I+1) \beta \hbar \boldsymbol{\omega}_i,
\qquad
\langle \hat{\bm{S}}_k \rangle = -\frac{1}{3} S(S+1) \beta \hbar \boldsymbol{\omega}_k.
\end{equation}
$I$ and $S$ are the spin quantum numbers for spins $I$ and $S$.
We then write a self-consistent set of equations, using the lattice Fourier
transforms of $\langle \hat{\bm{I}}_i \rangle$, $\langle \hat{\bm{S}}_k \rangle$, and
$\langle A_{ik} \rangle$:
\begin{equation}
\bigg\lbrace
\begin{array}{lclcl}
\lambda_I I^\alpha(\bm{k}) & + & c\gamma_I \gamma_S A^*(\bm{k}) S^\alpha (\bm{k})
& = & 0 \\
c\gamma_I \gamma_S A(\bm{k}) I^\alpha (\bm{k}) & + & \lambda_S S^\alpha (\bm{k})
& = & 0
\end{array}
\end{equation}
where $\lambda_I^{-1} = \beta I(I+1)/3 $, $\lambda_S^{-1} = \beta S(S+1)/3 $,
$\alpha \in \lbrace x,y,z \rbrace$, and $I^\alpha(\bm{k})$, $ S^\alpha (\bm{k})$, and
$A(\bm{k})$ are the lattice Fourier transforms of $\langle \hat{I}_i^\alpha \rangle$,
$\langle \hat{S}_k^\alpha \rangle$, and $\langle A_{ik} \rangle$ respectively.

We now calculate the transition temperature and ordered structure. The above
equations are satisfied for all $\bm{k}$ at the phase transition. Thermodynamic
stability \cite{Gold1974} implies that we should set all $I^\alpha(\bm{k})$, and
$S^\alpha(\bm{k})$ equal to zero, except the one which gives the maximum
$\left| T_c \right|$. We have to keep in mind that this can occur for small
$\left| \bm{k} \right|$, in which case the predicted ordered structure is
ferromagnetic with domains.

For a zincblende lattice, with coupling only between different species and the dc
magnetic field parallel to the $(111)$ crystalline direction, this maximum occurs for
$\bm{k}_0=a^{-1}(\frac{\pi}{2},\frac{\pi}{2},\frac{\pi}{2})$;
$\left| A(\bm{k}_0) \right|=7.00a^{-3}$, with $2a$ the edge of the conventional cubic
cell. The predicted ordered structure is the same as the one
described in Sec. \ref{sec:ExpSyst}, which followed from nearest neighbor
interactions alone. Because of the Heisenberg form of the interaction, there is
complete rotational degeneracy for classical spins. The transition temperature is
calculated to be
\begin{equation}
k_B T_c = \frac{\sqrt{I(I+1)S(S+1)}}{3} c\gamma_I\gamma_S \left| A(\bm{k}_0) \right|.
\end{equation}

The spin temperature of the system is modified by the ADRF process, and the subsequent
application of the pulse sequence. So, the relevant quantity is the critical entropy,
or the critical polarization in high field. To calculate it, we follow the formalism
described in Sec. \ref{sec:calculation}. To avoid complications, we use the high
temperature approximation. For a lattice with $N$ spins $I$ and $N$ spins $S$, the
entropy before ADRF is
\begin{eqnarray}
\frac{\mathcal{S}}{Nk_B} &=& \ln(2I+1) -\frac{1}{2} \beta_I^2\gamma_I^2 \hbar^2
\frac{I(I+1)}{3} \left( B_0\right)^2 + \nonumber \\
&+& \ln(2S+1) -\frac{1}{2} \beta_S^2\gamma_S^2 \hbar^2 \frac{S(S+1)}{3}
\left( B_0\right)^2,
\label{eq:SBefADRF}
\end{eqnarray}
while the polarization of spin-$\frac{1}{2}$ nuclei ($^{31}$P for InP) is
\begin{equation}
p = \tanh\left( \frac{\beta\gamma\hbar B_0}{2} \right).
\end{equation}
We first assume that both spin species always have the same spin temperature. This is
not true for the optical pumping mechanism for species with different spins and/or
different gyromagnetic ratios, and we study this effect later. During the application
of the pulse sequence on resonance to both species, the entropy is
\begin{multline}
\frac{\mathcal{S}}{Nk_B} = \ln(2I+1) + \ln(2S+1) \\
-\frac{1}{2} \beta^2 \frac{I(I+1)}{3} \frac{S(S+1)}{3}
(3c^2\gamma_I^2\gamma_S^2)\sum_k (A_{ik})^2.
\end{multline}
The above equation is valid only in the paramagnetic phase. For the zincblende
structure with the applied Zeeman magnetic field $\bm{B}_0\parallel (111)$,
$\sum_k (A_{ik})^2 = 13.238 a^{-6}$. At $\beta=\beta_c$, this expression for the
entropy is equal to the value in eq. (\ref{eq:SBefADRF}). If both species are
spin-$\frac{1}{2}$, and have identical gyromagnetic ratios, the critical polarization
is $p_c=56\%$.

The general case does not include any further complications. Just after ideal optical
pumping of the sample\cite{Dyakonov1974, Dyakonov1984}, the spin temperatures of the
two species satisfy
\begin{equation}
\beta_I \gamma_I = \beta_S \gamma_S.
\end{equation}
Equilibration between the two spin species occurs for
$B_I\sim\left( B_0 -\frac{\omega_I}{\gamma_I} \right)$,
$B_S\sim\left( B_0 -\frac{\omega_S}{\gamma_S} \right)$, and
$\gamma_I \left( B_0 -\frac{\omega_I}{\gamma_I} \right) \simeq \gamma_S
\left( B_0 -\frac{\omega_S}{\gamma_S} \right)$, where $B_I$, $B_S$ are the rf fields
acting on the $I$ and $S$ spins.
The first condition arises as we need the interaction term to not commute with the
rest of the Hamiltonian for $I$ and $S$ spins (see (\ref{eq:totHamiltonian})), so that
the two spin species can exchange energy.
The second condition maintains that simultaneous spin-flips conserve energy, so
that they have a non-negligible probability.
We assume that these fields are much larger than the local fields of Eq.
(\ref{eq:LocalField}), so that we can ignore the latter. We also assume that
equilibration happens at some particular point of ADRF, when the $I$ and $S$ spins
have temperatures $\beta_I^\prime$ and $\beta_S^\prime$ respectively. From equation
(\ref{eq:EntropyLocalField})
\begin{equation}
\begin{matrix}
\beta_I^\prime \left( B_0-\frac{\omega_I}{\gamma_I}\right) \sqrt{2} = \beta_I B_0 \\
\beta_S^\prime \left( B_0-\frac{\omega_S}{\gamma_S}\right) \sqrt{2} = \beta_S B_0
\end{matrix}
\bigg\rbrace \Rightarrow \frac{\beta_I^\prime}{\beta_S^\prime} \simeq
 \frac{\beta_I \gamma_I}{\beta_S \gamma_S}=1.
\end{equation}
That is, the two species are already in equilibrium, and there is no entropy loss.

It is now straightforward to calculate the critical polarization for both species,
with the assumption of ideal optical pumping and no entropy loss during the
transformation of the Hamiltonian. The result is
\begin{equation}
p_c^{\rm P} = 15\%, \qquad p_c^{\rm In} = 49\%.
\end{equation}

\section*{Conclusion}
We have proposed an experiment of quantum simulation of spin Hamiltonians by the
manipulation of nuclear spins in a solid. Bulk InP seems to be a suitable material for
this experiment. A particular orientation of the crystal produces a convenient
model Hamiltonian. We have presented an NMR pulse sequence, which transforms the 
natural dipolar spin Hamiltonian to a Heisenberg model with long range interactions.
We have also performed a calculation to estimate the experimental feasibility of our
proposal. Our results suggest that initial polarization of Phosphorous nuclei of about
15\%, and of Indium nuclei of about 50\% is needed in InP, in order for a phase
transition to be observed. The difference in these values is because of different
nuclear spin quantum numbers, and this values should be consistent under ideal
optical pumping conditions. This critical polarization seems reasonable for optical
pumping of bulk InP in low temperature and high magnetic field.

This scheme provides an alternative for quantum simulation experiments which is
predicted to be technically simpler than the cold atom approach, and involves a
macroscopic number of quantum particles.
It is true that its flexibility in implementing a specific spin-lattice Hamiltonian is limited, 
and the method of detection described in the introduction cannot give any information
about the specific orientation of the spin dipoles. To extract this information, one needs
more complicated techniques, such as neutron scattering \cite{Oja1997}. More
sophisticated NMR measurement strategies might also be possible.
Nevertheless, this scheme enables the study of several interesting spin-lattice models.
Similar experiments have already been performed \cite{Chapellier,Quiroga1,Quiroga2},
and they constitute a proof of principle for the validity of the spin-temperature concept,
and for this method of manipulating the dipolar interaction.
However, in these experiments only one spin species was used, which limits the freedom to
manipulate the internal Hamiltonian. In particular,
$\frac{\pi}{2}$-pulses could not be used, as the interaction Hamiltonian is averaged
out, and the equilibration time becomes infinite. In our case, the interaction between
different spin species remains non-negligible, and the adiabatic criterion can be
satisfied near resonance, when the two spin species can exchange energy.

Our proposal is known to have several uncertainties and approximations, discussed
below. To our knowledge, nuclear ordering has not been observed until now using optical
pumping as the cooling technique. This renders the experiment both interesting and
technically challenging. We also have to point out that this experiment, like most NMR
experiments, involves subtleties because of various relevant timescales, and the
inequalities that they have to satisfy. A detailed discussion is quite technical, but
plenty of information can be found in the literature, in particular about when AHT is
valid \cite{Haeberlen}, and when ADRF \cite{SlichterBook} is adiabatic.
What is not adequately studied, however, and requires further research, is the rate of
establishment of thermal equilibrium in low spin temperature while a multipulse
sequence is applied.
This scheme also suffers from several non-idealities, namely indirect nuclear magnetic
interactions \cite{Tomaselli1998,Iijima2006}, non-idealities of the multipulse sequence,
or  not perfectly adiabatic transformation of the Hamiltonian. Nonetheless, non-ideal
ADRF is found in Ref. \onlinecite{Gold1968} not to pose a serious problem for high
initial polarization and typical NMR applied magnetic fields. Further, the method
described for the calculation of the transition temperature is valid only for classical
spins within the mean-field approach, so a more sophisticated calculation would be
interesting.

\begin{acknowledgments}
We thank Thaddeus D. Ladd for helpful discussions.
Financial support was provided by JST SORST program on Quantum Entanglement.
\end{acknowledgments}

\appendix

\section{\label{sec:GeneralCase}The semi-invariants for the general dipolar interaction case}
In Sec. \ref{sec:LongitudinalCase}, we calculated the semi-invariants in
the case that the spin interaction is longitudinal. We now restore the
transverse term of the Hamiltonian
\begin{equation}
\mathcal{\hat{H}} = \Delta \sum_i \hat{I}_i^z 
+ \frac{1}{2}\sum_{i,j}\left( u_{ij}\hat{I}_i^z \hat{I}_j^z +v_{ij}\hat{I}_i^+\hat{I}_j^- \right)
\equiv \mathcal{\hat{H}}_Z + \mathcal{\hat{H}}^\prime_D.
\end{equation}
For the dipolar interaction, $v_{ij}=-\frac{1}{2}u_{ij}$. We take into
account that the operators $\hat{I}_i^z,\hat{I}_i^\pm$ do not commute and introduce
time-ordering. Details can be found in Refs.~\onlinecite{SHEB62,AbrGold82}.
The semi-invariants are defined in a similar way as in (\ref{eq:SemiInvDef}), but
the partition sets of the right part must follow the ordering of the left hand side.
The additional semi-invariants we need for our calculation are:
\begin{subequations}
\begin{align}
M^0_2(I^+I^-) & = j(j+1) + t_j^{(1)} - t_j^{(2)},\\
M^0_2(I^-I^+) & = j(j+1) - t_j^{(1)} - t_j^{(2)}, \\
M^0_3(I^zI^+I^-) &= M^0_3(I^+I^-I^z) =  \nonumber \\
- \left(t_j^{(1)}\right)^2 &+ t_j^{(2)} \left(1+t_j^{(1)}\right) - t_j^{(3)},\\
M^0_3(I^zI^-I^+) &= M^0_3(I^-I^+I^z) = \nonumber \\
\left(t_j^{(1)}\right)^2  &- t_j^{(2)} \left(1-t_j^{(1)}\right) - t_j^{(3)},\\
M^0_3(I^+I^zI^-) &+ M^0_1 M^0_2(I^+I^-) = \nonumber \\
-j(j+1) &+ (j^2+j-1)t_j^{(1)} + 2 t_j^{(2)} - t_j^{(3)},\\
M^0_3(I^-I^zI^+) &+ M^0_1 M^0_2(I^-I^+) = \nonumber \\
j(j+1) &+ (j^2+j-1)t_j^{(1)}  - 2 t_j^{(2)} - t_j^{(3)}.
\end{align}
\end{subequations}
The functions $t_j^{(n)}$ are defined in (\ref{eq:fjDef}).

\section{\label{sec:CalcDipolarInt}Calculation of the dipolar interaction energy}
\begin{figure}[thb]
\includegraphics[width=3.2in]{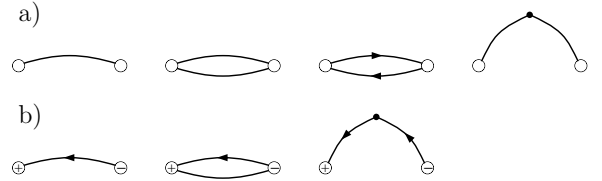}
\caption{\label{fig:diagrams}Diagrams for the calculation of the expansions in
equations a) (\ref{eq:IzIz}) and b) (\ref{eq:I+I-}). The diagrammatic rules are presented
in Refs.~\onlinecite{AbrGold82,SHEB62}.}
\end{figure}

As is clear in Sec. \ref{sec:CalcThermQuant}, calculation of the dipolar
interaction energy is the key to the calculation of other interesting
thermodynamic quantities, and in particular the entropy, which is our ultimate goal.
The dipolar energy is
\begin{equation}
\frac{2}{N} \langle \mathcal{\hat{H}}^\prime_D \rangle = \sum_i \left[ u_{1i}
\langle \hat{I}_1^z\hat{I}_i^z\rangle + \frac{v_{1i}}{2}
\left( \langle \hat{I}_1^+\hat{I}_i^-\rangle+ \langle \hat{I}_1^-\hat{I}_i^+\rangle \right) \right],
\end{equation}
with $v_{ij} = -\frac{u_{ij}}{2}$. $N$ is the total number of spins. Because of the
symmetry of the resulting diagrams, the terms $\langle \hat{I}_1^+\hat{I}_i^-\rangle$ and
$\langle \hat{I}_1^-\hat{I}_i^+\rangle$ give the same contribution, so we keep just the former,
multiplied by $v_{1i}$. The second order expansions give the diagrams of
Fig.~\ref{fig:diagrams} and result in the following expressions

\begin{align}
\label{eq:IzIz}
\langle \hat{I}^z_1 \hat{I}^z_i \rangle &\simeq
(-\beta)(M^0_2)^2u_{1i}+ (-\beta)^2\frac{1}{2}(M^0_3)^2(u_{1i})^2 + \nonumber \\ 
+&(-\beta)^2 M^0_3(+-z)M^0_3(-+z)(\frac{v_{1i}}{2})^2 + \\
+& (-\beta)^2 (M^0_2)^3 \sum_j u_{1j} u_{ji}, \nonumber 
\end{align}
\begin{widetext}
\begin{align}
\label{eq:I+I-}
\langle \hat{I}^+_1 \hat{I}^-_i\rangle &\simeq (-\beta)\frac{v_{1i}}{2} M^0_2(+-)M^0_2(-+)+
(-\beta)^2 \frac{v_{1i}}{2} u_{1i} \frac{1}{2}
[M^0_3(+z-)M^0_3(-z+)+M^0_3(z+-)M^0_3(z-+)] + \nonumber \\
+& (-\beta)^2 \sum_j \frac{v_{1j}}{2} \frac{v_{ji}}{2} M^0_2(+-) M^0_2(-+)
\frac{1}{2}[M^0_2(+-)+M^0_2(-+)].
\end{align}
\end{widetext}
The final result can be expressed in terms of the following quantities,
which depend on the particular lattice and the orientation of the magnetic field:
\begin{subequations}
\begin{align}
I_2 & = \sum_i(u_{1i})^2, \\
I_3 & = \sum_i(u_{1i})^3, \\
K_3 & = \sum_{i,j}u_{1i}u_{ij}u_{j1}.
\end{align}
\end{subequations}
For the case of an fcc lattice, with the magnetic field parallel to the (111)
crystalline direction, we obtain the values
\begin{subequations}
\begin{align}
I_2 & = 1.6908 \left( \frac{\mu_0}{4\pi} \frac{\hbar^2 \gamma^2}{a^3} \right)^2,
\end{align}
\begin{align}
I_3 & = -0.0072073 \left( \frac{\mu_0}{4\pi} \frac{\hbar^2 \gamma^2}{a^3} \right)^3, \\
K_3 & = 2.1173 \left( \frac{\mu_0}{4\pi} \frac{\hbar^2 \gamma^2}{a^3} \right)^3,
\end{align}
\end{subequations}
where $2a$ is the edge of the conventional cubic cell.
We refer to these values in Sec. \ref{sec:results}.

With these definitions, we have the following expansion for the dipolar interaction
energy:
\begin{multline}
\frac{1}{N} \langle \mathcal{\hat{H}}^\prime_D \rangle =
\beta G_1(t^{(1)}(\eta),t^{(2)}(\eta)) + \\ 
\beta^2 G_2(t^{(1)}(\eta),t^{(2)}(\eta),t^{(3)}(\eta)) + \ldots ;
\end{multline}
for $v_{ij}=-\frac{1}{2}u_{ij}$, the functions $G_1$, and $G_2$ take the form
\begin{widetext}
\begin{subequations}
\begin{equation}
G_1\left(t^{(1)}(\eta),t^{(2)}(\eta)\right) =
-\frac{1}{2} \left[ (M^0_2)^2 I_2 + \frac{1}{8} M^0_2(+-)M^0_2(-+)I_2 \right]
\end{equation}
\begin{eqnarray}
G_2 \left(t^{(1)}(\eta),t^{(2)}(\eta),t^{(3)}(\eta)\right) =
\frac{1}{2} \Bigl[ \frac{1}{2} (M^0_3)^2I_3 + \frac{1}{16}M^0_3(+-z)M^0_3(-+z)I_3 
+ (M^0_2)^3K_3 +\nonumber \\
+ \frac{1}{16} \left(M^0_3(+z-)M^0_3(-z+)+M^0_3(z+-)M^0_3(z-+)\right) I_3
-\frac{1}{64} M^0_2(+-)M^0_2(-+)\left(M^0_2(+-)+M^0_2(-+)\right) K_3 \Bigr].
\end{eqnarray}
\end{subequations}
\end{widetext}

\end{document}